\documentclass[10pt,a4paper]{article}
\newif\ifjhepstyle
\IfFileExists{jheppub.sty}{\jhepstyletrue\usepackage{jheppub}}{\jhepstylefalse}
\usepackage[T1]{fontenc}
\usepackage{lmodern,amsmath,array,booktabs,multicol,microtype,xcolor}
\ifjhepstyle\else
\usepackage[margin=22mm]{geometry}
\usepackage[colorlinks=true,linkcolor=blue!50!black,urlcolor=blue!50!black,citecolor=blue!50!black]{hyperref}
\fi
\hypersetup{pdftitle={Gotta Catch 'Em All: Uncovered LHC signatures},pdfauthor={Jeremi Niedziela}}
\newcommand{\appheading}[1]{\par\medskip\textbf{#1}\par\smallskip}
\newcommand{\appfamily}[1]{\par\smallskip\textbf{#1}\par\nopagebreak}
\newcommand{\papertitle}{Gotta Catch 'Em All: Uncovered LHC signatures}
\newcommand{\paperauthor}{Jeremi Niedziela}
\newcommand{\paperaddress}{Inter-University Institute for High Energies (IIHE), Vrije Universiteit Brussel,\newline Pleinlaan 2, 1050 Brussels, Belgium}
\newcommand{\paperemail}{jeremi.niedziela@cern.ch}
\newcommand{\paperabstract}{
  
Each combination of physics objects defines a signature in which new physics can be sought at the LHC. This paper attempts to quantify how many of these signatures have been covered. I~compare combinations of electrons, muons, tau leptons, photons, jets, and heavy Standard Model particles with a catalogue based on public ATLAS, CMS, and LHCb results. Matching is exact in the visible objects, except that a $b$ jet can also fill a generic jet slot, and it includes an object-content proxy for CMS MUSiC. No exact match is found in this catalogue for 4 of 55 two-object, 59 of 220 three-object, and 507 of 715 four-object signatures. Of these 570 signatures, 69 can arise in simplified beyond-the-Standard-Model (BSM) scenarios. The census identifies promising candidates for dedicated searches and reinterpretations.

}

\ifjhepstyle
\notoc
\title{\papertitle}
\author{\paperauthor}
\affiliation{\paperaddress}
\emailAdd{\paperemail}
\abstract{\paperabstract}
\else
\title{\papertitle}
\author{\paperauthor\\[3pt]
\small Inter-University Institute for High Energies (IIHE), Vrije Universiteit Brussel\\[-1pt]
\small Pleinlaan 2, 1050 Brussels, Belgium\\[-1pt]
\small \texttt{\paperemail}}
\date{}
\fi
\begin{document}
\maketitle
\ifjhepstyle\else
\begin{abstract}
\paperabstract
\end{abstract}
\fi

\section{Introduction}
LHC searches test a wide range of models. Here I ask a simple question: which combinations of physics objects are explicitly targeted by existing searches? The Les Houches report organized prospective searches by final-state objects~\cite{lh}. Fischer et al. discussed how signals can escape existing searches, for example through soft objects or associated production~\cite{hidden}. SModelS identifies missing simplified topologies for a given spectrum~\cite{smodels}, while surveys of exotic Higgs decays organize possible searches within one sector~\cite{curtin}. CMS MUSiC and the ATLAS general search tested many reconstructed event classes directly against data~\cite{music,atlasgeneral}. Closest to the present study, Craig et al. surveyed searches for resonances decaying into pairs of visible particles and identified unexplored combinations~\cite{twobody}. Kim et al. later reviewed their motivation and status and found significant gaps, particularly where one or both decay products are not Standard Model particles~\cite{twobodystatus}. Here I consider combinations of up to six familiar objects, without requiring them to come from a single resonance, and compare them with a catalogue of published results.

I enumerate the combinations of a set of familiar objects and compare them with published searches and other results. This produces a long list of candidate gaps. To see which could be worth following up, I examine possible production and decay processes.

\section{Definition of a signature and search coverage}
\label{sec:method}
A signature is an unordered multiset of $\{e,\mu,\tau,\gamma,j,b,t,W,Z,H\}$. A $j$ is a jet with no flavour requirement, including jets from gluons and $b$ quarks. A $b$ denotes a $b$-tagged jet or a specified bottom-quark jet in a signal topology. A heavy particle counts as one object if a search either tags it directly or identifies it as the parent of specified decay products. Here, $H$ denotes the 125 GeV Higgs boson. The labels $W$ and $Z$ also include off-shell bosons when the search explicitly identifies them in the decay chain. The signatures $HHH$ and $bbbbbb$ can describe the same search at different levels. The number of signatures with exactly $n$ objects is $\binom{n+9}{9}$. The main census ignores charge and the presence or absence of a missing transverse momentum (MET) requirement.

\paragraph{Publications and attribution.}
The main census uses public results available by 26 November 2026 and applies the same signature-matching rules to all three experiments. I reviewed the titles and available abstracts of 1,617 CMS, 1,517 ATLAS, and 885 LHCb publications and consulted the full papers where needed to identify the targeted channels. The catalogue contains entries from 880 CMS results (EXO, B2G, SUS, NPS, HIG, TOP, SMP, QCD, BPH, EWK, FSQ, FWD, and HIN), 793 ATLAS results (EXOT, HDBS, HIGG, HIGP, HMBS, STDM, SUSY, TOPQ, BPHY, and HION), and 19 LHCb results (resonances, Higgs production, lepton-flavour violation, and long-lived particles). The catalogue includes proton and ion collisions and entries with open multiplicities. Only entries with a fixed object count contribute to the exact census. Several results can support the same signature. A search restricted to the decay of an identified hadron is outside the scope of this census. For example, a limit on $B_s^0\to\mu^+\mu^-$ applies to a specific meson decay, whereas a direct $A'\to\mu^+\mu^-$ resonance search contributes to the dimuon signature. Most of the 885 LHCb publications concern identified hadrons, which largely explains why only 19 have catalogue entries. The publication records come from the official CMS and ATLAS lists, with INSPIRE used in addition for CMS and as the source of the LHCb records. The collaborations maintain full publication lists for \href{https://cms-results.web.cern.ch/cms-results/public-results/publications/}{CMS}, \href{https://atlaspo.cern.ch/public/twiki/publications/publications.html}{ATLAS}, and \href{https://cds.cern.ch/collection/LHCb%20Papers?ln=en}{LHCb}. Standard Model measurements are included only if they also provide a BSM interpretation.

\paragraph{Matching.}
I match each signature to the full set of visible objects in a documented signal topology or analysis category. An identified Standard Model parent can count as one object or be replaced by its specified visible daughters. The other branches and any visible production partners stay in the signature. Different branches of the same documented topology can be represented at different levels, even if the source does not name that mixed combination, as long as it specifies each decay being resolved. Each branch is counted once. For analysis categories, I use the documented object counts and identifications. A $b$ can fill a generic $j$ slot at the same multiplicity, so I also count signatures obtained by ignoring any number of $b$ tags. Table~\ref{tab:matching_examples} illustrates these rules.

\begin{table}[!htbp]
\centering
\small
\begin{tabular}{@{}p{0.48\linewidth}ll@{}}
\toprule
Search or category & Matches & Does not match\\
\midrule
Diboson search, $WZ$ with $W\to e\nu$~\cite{diboson} & $WZ$, $eZ$ & $e$ alone\\
Scalar leptoquark pair search, $LQ\to te$~\cite{atlaslq} & $ttee$ & $te$\\
$tW$ resonance search, leptonic $W$~\cite{cmstw} & $tW$, $te$, $t\mu$ & $t$ alone\\
ATLAS general search, $3b1j$ class~\cite{atlasgeneral} & $bbbj$, $bjjj$ & $bbbb$\\
\bottomrule
\end{tabular}
\caption{Example matches for the cited topology or category. Other searches can match the signatures in the last column.}
\label{tab:matching_examples}
\end{table}

I leave multiplicities given only as lower bounds open. The match concerns the documented objects, so additional radiation and requirements on missing transverse momentum can affect acceptance. Likewise, a match obtained by ignoring a $b$ tag does not establish sensitivity to every jet flavour. In signal models with an associated object, such as a $W$ or a quark accompanying three top quarks~\cite{threetop}, that object remains part of the topology.

For CMS MUSiC, the census adds an object-content proxy for classes with at least one $e$ or $\mu$ and any combination of $e,\mu,\gamma,j,b$~\cite{music}. An eligible class may still have been absent from the published scan. In particular, MUSiC merges classes with at least five non-$b$ jets into inclusive bins. A generic $j$ in this proxy can also stand for a jet from a lower-$j$ class whose $b$ tag is ignored. The ATLAS general search reported yields for 704 classes and retained 686 for the statistical analysis~\cite{atlasgeneral}. I transcribe nine distinct classes with documented scan results. Five come from Tables~4 and~5, which list the three smallest channel $p$-values for each of two distributions. Four more come from the example scan regions in Fig.~4. A class that appears only in a yield plot, such as the $\gamma\gamma b$ class, may have been scanned, but the plot does not establish whether the class had a valid region. Here \emph{uncovered} means unmatched in this catalogue under the stated rules. A search for a larger topology or an inclusive search may still be relevant to such a signature.

\section{Results}
\label{sec:results}
Table~\ref{tab:coverage} assigns each signature to the first row whose requirement it satisfies. Objects in the \emph{Also allowed} column are optional. The \emph{Earlier rows} column lists objects that would place the signature in a row higher in the table. For example, a signature with both $H$ and $t$ goes in the $H$ row. This classification does not imply any event veto. Among the 990 signatures with two to four objects, 570 (57.6\%) are unmatched. All ten single-object signatures have a match. All four unmatched two-object signatures, and 26 of the 59 three-object ones, are contained in a matched signature with one additional object. They are gaps in exact matching, not necessarily objects absent from all searches. Without the CMS MUSiC proxy, the numbers of unmatched signatures with two through six objects are 4, 59, 540, 1,915, and 4,886.

\begin{table}[!htbp]
\centering
\small
\setlength{\tabcolsep}{4pt}
\begin{tabular}{@{}lllr r r r r r@{}}
\toprule
& & & \multicolumn{5}{c}{Exact number of objects} & \\
\cmidrule(lr){4-8}
Required & Also allowed & Earlier rows & 2 & 3 & 4 & 5 & 6 & Total 2--6\\
\midrule
$H$ & $t,V,\ell,\gamma,j,b$ & --- & 1 & 24 & 183 & 706 & 2{,}002 & 2{,}916\\
$t$ & $V,\ell,\gamma,j,b$ & $H$ & 0 & 12 & 129 & 478 & 1{,}250 & 1{,}869\\
$W$ or $Z$ & $\ell,\gamma,j,b$ & $H,t$ & 2 & 18 & 154 & 520 & 1{,}237 & 1{,}931\\
$\tau$ & $e,\mu,\gamma,j,b$ & $H,t,V$ & 1 & 4 & 36 & 119 & 232 & 392\\
$e$ or $\mu$ & $\gamma,j,b$ & $H,t,V,\tau$ & 0 & 0 & 0 & 0 & 0 & 0\\
$\gamma$ & $j,b$ & $H,t,V,\ell$ & 0 & 1 & 5 & 12 & 20 & 38\\
$j$ or $b$ & --- & $H,t,V,\ell,\gamma$ & 0 & 0 & 0 & 6 & 0 & 6\\
\midrule
Total &  &  & 4 & 59 & 507 & 1{,}841 & 4{,}741 & 7{,}152\\
Uncovered fraction & & & 7.3\% & 26.8\% & 70.9\% & 92.0\% & 94.7\% & 89.4\%\\
\bottomrule
\end{tabular}
\caption{Counts of uncovered signatures. $V$ denotes $W$ or $Z$ and $\ell$ denotes $e$, $\mu$, or $\tau$. On a row with multiple required alternatives, one or more of them may occur. Further copies of a required object are allowed. The columns contain 55, 220, 715, 2,002, and 5,005 possible signatures, respectively, for a combined denominator of 7,997.}
\label{tab:coverage}
\end{table}

I include five- and six-object signatures for completeness. The model discussion and the appendix focus on signatures with up to four objects. At higher multiplicities, analyses often leave some object counts open. The five- and six-object fractions therefore describe gaps in the exact catalogue encoding, although inclusive searches may accept events in these categories.

\pagebreak[4]
\paragraph{Effect of conservation laws on coverage.}
In a separate check, which includes single-object signatures, I count each signature twice: with and without a recorded MET requirement. A catalogue entry matches only its own MET category, except that the MUSiC proxy is used in both. Entries with an unset MET flag are treated as having no recorded requirement. In the MET category, I allow extra neutrinos to balance lepton number. In the other category, the charged-lepton count must be even, either in total or within each flavour. Once these lepton requirements are imposed, electric charge, baryon number, and fermion parity remove no further entries. Here particle and antiparticle assignments are free, and each of the two initial partons can be a quark, an antiquark, or a gluon. For every combination of up to four objects, the assignments can then be chosen so that the electric charge and baryon number agree with those of an initial state. Since baryon number also fixes whether the numbers of quarks in the initial and final states differ by an even number, fermion parity only requires an even total number of leptons, including neutrinos. This is ensured by the even charged-lepton count without MET and can be arranged with additional neutrinos in the MET category.

Splitting by MET raises the unmatched fraction even before any conservation requirement is applied. For one to three objects it is $241/570$ (42.3\%), compared with $63/285$ (22.1\%) without the split. For one to four objects it is $1{,}344/2{,}000$ (67.2\%), compared with $570/1{,}000$ (57.0\%). This difference arises because each signature must be matched separately in the two MET categories. The effect of the conservation requirements is measured relative to the split samples. With total lepton number conserved, the results are $181/452$ (40.0\%) for up to three objects and $994/1{,}560$ (63.7\%) for up to four. With each lepton flavour conserved separately, they are $172/428$ (40.2\%) and $930/1{,}443$ (64.4\%). The unmatched fraction therefore changes by less than three and a half percentage points when these requirements are imposed.

\section{Possible signals and indirect sensitivity}
\label{sec:models}
For each unmatched signature with up to four visible objects, I ask whether a decay or a production and decay process in a simplified BSM model gives exactly that combination of visible objects. Invisible particles are allowed and no minimum production rate is required. Some matches specify only a decay, so a viable production mechanism would still need to be established. If several models match, the appendix lists one representative model family.

\begin{description}
\setlength{\itemsep}{1pt}
\item[\textnormal{\textbf{S, C}}] Neutral scalars and axion-like particles, or charged and top-associated resonances.
\item[\textnormal{\textbf{L, LQ}}] Heavy leptons and leptoquarks, with decays such as $E\to\tau H$, $N\to\tau W$, and $\mathrm{LQ}\to t\tau$~\cite{vlreview}.
\item[\textnormal{\textbf{F, E}}] Excited fermions and supersymmetric cascades. For example, an excited $\tau$ can decay to $\tau\gamma$.
\end{description}

\begin{table}[!htbp]
\centering
\begin{tabular}{lrrr}
\toprule
Exact multiplicity & Uncovered & With a model match & No model match found\\
\midrule
2 & 4 & 4 (100.0\%) & 0 (0.0\%) \\
3 & 59 & 20 (33.9\%) & 39 (66.1\%) \\
4 & 507 & 45 (8.9\%) & 462 (91.1\%) \\
\midrule
Total & 570 & 69 (12.1\%) & 501 (87.9\%) \\
\bottomrule
\end{tabular}
\caption{Uncovered signatures with up to four objects and their matches to simplified BSM models. The fractions depend on the models and decay modes considered.}
\label{tab:models}
\end{table}

Two examples show how a signal model can differ from the topology targeted by a related analysis. A $\tau\gamma$ resonance could arise from charged-current production $q\bar q'\to W^*\to\tau^*\nu_\tau$, followed by $\tau^*\to\tau\gamma$. The CMS excited-$\tau$ search targets $\tau^*\tau$ and fits $\tau_{\mathrm h}\tau_{\mathrm h}\gamma$, $e\tau_{\mathrm h}\gamma$, and $\mu\tau_{\mathrm h}\gamma$~\cite{taustar}. Each category has a visible object from the second $\tau$ decay. At higher multiplicity, a heavy resonance decaying to two scalars, each decaying to $HH$, gives $HHHH$. A four-Higgs signal can also enter a triple-Higgs selection, as shown in a dedicated phenomenological study~\cite{fourh}.

A search for one signature can also accept events from another. For signatures with up to four objects, I check for possible overlap from a single leptonic $\tau$ decay by replacing one $\tau$ by an $e$ or $\mu$ and leaving the other objects fixed. This finds a match for 50 of 201 unmatched signatures containing a $\tau$ ($24.9\%$). Of those, 31 match an explicit entry and 19 depend on the MUSiC proxy alone. These numbers count only possible overlaps. Branching fractions, daughter momenta, neutrinos, and selection cuts determine whether those overlaps give useful sensitivity. Fiducial measurements can also constrain BSM signals through reinterpretation, for example with Contur~\cite{contur}.

\paragraph{Displaced objects.}
I extend the two- to four-object census by giving each of the ten object types a displaced counterpart $X_{\mathrm d}$. The label means that $X$ originates at a displaced decay. For photons and heavy Standard Model particles such as $H$, $Z$, or $t$, it refers to the production point, even if a heavy particle then decays promptly. The original signatures have no displacement requirement. In the new signatures, any combination of one or more objects can carry the displaced label. For this extension I use a separate list of published topologies and analysis categories that specify which objects are displaced. I apply the same $b$-to-$j$ rule to these entries. The assignments can overlap in data.

With twenty labels, there are 210, 1,540, and 8,855 exact two-, three-, and four-object signatures. The total of 10,605, which includes 9,615 variants with at least one displaced object, is 10.7 times the original 990. The single displacement label involves no subdivision by decay length, vertex multiplicity, or which objects share a vertex.

A targeted scan of CMS, ATLAS, and LHCb results documents 36 of the added signatures. Examples include displaced dileptons and heavy neutral leptons~\cite{cmsdilep,atlasdilep,lhcbdimuon,atlashnl,cmshnlj}, dijets and four-fermion topologies from long-lived scalars~\cite{lhcbdijet,cmsdj}, photons from two displaced decays~\cite{atlasdphoton}, and pairs of $H$ or $Z$ bosons produced in displaced neutralino decays~\cite{cmsdHZ}. A match can be a stated signal topology whose displacement is probed through its daughters. These 36 matches are counted after applying the $b$-to-$j$ rule. Including them, $10{,}149/10{,}605$ (95.7\%) signatures are unmatched in the expanded census. This fraction should not be compared directly with the baseline result, because the displaced matches come from a much more limited targeted scan. The main lesson is how quickly the space of signatures, and with it the potential unexplored territory, grows when one additional property is distinguished. The scan includes a preliminary CMS result~\cite{cmshnlpas}.

\section{Most promising unmatched signatures}
\label{sec:promising}
The signatures shown in red in the appendix have a match to at least one of the models considered. Among them, I highlight combinations with few objects, a comparatively simple production or decay chain, and intermediate states that could be reconstructed. This is a qualitative selection. I have not compared cross sections, branching fractions, existing limits, or expected sensitivities. Table~\ref{tab:promising} gives one example process for each choice and a related search or measurement.

\begin{table}[!htbp]
\centering
\small
\setlength{\tabcolsep}{4pt}
\renewcommand{\arraystretch}{1.15}
\begin{tabular}{@{}>{\raggedright\arraybackslash}p{0.13\linewidth}>{\raggedright\arraybackslash}p{0.34\linewidth}>{\raggedright\arraybackslash}p{0.47\linewidth}@{}}
\toprule
Signature & Example process & Related result and missing element\\
\midrule
$\tau H$, $\tau Z$, $\tau W$ & Single production of a heavy lepton with a neutrino: $W^*\to L\nu_\tau$ with $L\to\tau H$ or $\tau Z$, or $Z^*\to N\bar\nu_\tau$ with $N\to\tau W$ & The vector-like lepton pair search keeps both decay branches~\cite{atlasvll}. A category with one $\tau$ and one boson is missing.\\
$\gamma\tau$ & $W^*\to\tau^*\nu_\tau$, $\tau^*\to\tau\gamma$ & The excited-$\tau$ search has an additional $\tau$ decay product~\cite{taustar}. A single-$\tau$ plus photon category is missing.\\
$\gamma ZZ$ & $q\bar q\to Z^*\to Za$, $a\to Z\gamma$, where $a$ is an axion-like particle & $ZZ\gamma$ production has been measured~\cite{atlaszzg,cmszzg}. A search for the $Z\gamma$ resonance, or a reinterpretation, is missing from the catalogue.\\
$bZZ$, $bHH$ & $gb\to bS$, with a neutral scalar $S\to ZZ$ or $HH$ & A related $S\to ZZ$ search targets gluon and vector-boson fusion~\cite{cmszz}. A dedicated category with an associated $b$ jet is missing.\\
$\gamma\gamma b$ & $gb\to bS$, $S\to\gamma\gamma$ & Inclusive diphoton searches can accept such events~\cite{cmsdiphoton}. A dedicated $b$-tagged resonance category is missing from the catalogue.\\
$WZH$ & $q\bar q'\to X^\pm\to W^\pm A$, $A\to ZH$ & Diboson searches reconstruct two-body resonances~\cite{diboson}. A search for the full three-boson cascade is missing.\\
\midrule
$HHHH$, $ZHHH$ & $X\to S_1S_2$, with $S_{1,2}\to HH$, or $S_1\to ZH$ and $S_2\to HH$ & Triple-Higgs searches have been performed~\cite{atlashhh,cmshhh}. Selections reconstructing a fourth boson are missing.\\
\bottomrule
\end{tabular}
\caption{Selected unmatched signatures with a model match. The processes are examples whose rates and viability require dedicated study. The statements about missing categories refer to the catalogue and matching rules used here.}
\label{tab:promising}
\end{table}

The two-object cases are the simplest. For $\tau H$, $\tau Z$, and $\tau W$, a single heavy lepton could give a resonance in the visible $\tau$ and boson system, although neutrinos from the $\tau$ decay, and for some boson decays, limit the mass resolution. The production rate depends on the mixing of the heavy lepton with the $\tau$ lepton and could be small. For $\gamma\tau$, leptonic $\tau$ decays lead to an electron or muon, a photon, and MET. Existing selections in these final states could therefore have some sensitivity.

Among the three-object cases, $\gamma ZZ$ is a natural starting point for reinterpretation, since measurements of the same three bosons exist. The $b$-associated processes are relevant for new scalars with enhanced couplings to $b$ quarks. Their value relative to inclusive searches depends on this production mechanism and on the gain from requiring a $b$ jet. For $\gamma\gamma b$, the diphoton branching fraction is an additional consideration. I do not highlight $jHH$ or $jZZ$, since an extra jet can arise from initial-state radiation and be accepted by inclusive two-boson searches.

The four-object examples $HHHH$ and $ZHHH$ offer several intermediate mass constraints, at the cost of small branching fractions for specific final states and large combinatorics. A four-Higgs signal can also enter a triple-Higgs selection, as studied phenomenologically~\cite{fourh}.

Some other signatures in red need more specific assumptions. For example, the leptoquark processes considered for $\tau bb$, $\tau jb$, and $\tau tt$ combine different decay branches that, for some combinations, require leptoquark states with different electric charges. A related search selects one hadronically decaying $\tau$, at least two $b$ jets, and MET~\cite{atlastaub}. It is not counted as a match for $\tau bb$ because lower-bound multiplicities are left open, but it may already be sensitive to such signals. Several supersymmetric cascades similarly require a particular set of particles and couplings.

\section{Outlook}
The most immediate targets identified here are resonances in $\tau$ plus $H$, $Z$, $W$, or $\gamma$, together with selected multiboson and $b$-associated signatures. Existing measurements and nearby selections, such as those for $ZZ\gamma$ production or inclusive diphoton resonances, are a natural place to start. Estimates of production rates and sensitivity are needed to decide which of these gaps deserve a dedicated search.

This census has important limitations. It does not assess the acceptance of existing analyses, so an unmatched signature may already be constrained by a larger, nearby, or inclusive selection. The catalogue can also miss subchannels of individual publications. In particular, I have not systematically expanded every mixed parent and daughter representation allowed by the matching rule, including those with four or more objects. Some signatures listed as unmatched may therefore have such a match. The model matches depend on the processes considered and do not establish that a signal is viable or observable.

Across all exact combinations with two to six visible objects, 7,152 of 7,997 signatures are unmatched. This total is dominated by five- and six-object combinations. The enumeration could be extended to distinguish MET, displaced or delayed objects, low $p_T$, or particular charge combinations. Such distinctions matter even for matched signatures. For instance, a prompt high-$p_T$ selection and a low-$p_T$ displaced selection can probe very different signals with the same visible particles.


\appendix
\section{Uncovered signatures with up to four objects}
Entries use the object conventions of Sec.~\ref{sec:method}. A repeated letter denotes a distinct object, as in $HHHH$. Here $V\in\{W,Z\}$ and $\ell\in\{e,\mu,\tau\}$. Each occurrence is expanded independently, including in $VV$ or $\ell\ell$. I use this shorthand only when all expansions are unmatched and share the same model-family label. The union of the expansions contains all 570 signatures, with each appearing in exactly one entry. A superscript identifies one model family from Sec.~\ref{sec:models}. Entries with a model match are shown in red. An asterisk marks a signature with no match among the models considered. A red entry is not necessarily among the selected cases of Sec.~\ref{sec:promising}.

\begin{multicols}{2}
\raggedright\small
\appheading{Exactly 2 visible objects (4 signatures)}
\mbox{\textcolor{red}{\mbox{$\tau{}H{}$\textsuperscript{L}}}, \textcolor{red}{\mbox{$\tau{}W{}$\textsuperscript{L}}}, \textcolor{red}{\mbox{$\tau{}Z{}$\textsuperscript{L}}}, \textcolor{red}{\mbox{$\gamma{}\tau{}$\textsuperscript{F}}}}.\par
\appheading{Exactly 3 visible objects (59 signatures)}
\appfamily{Higgs (24)}
\textcolor{red}{\mbox{$\gamma{}\ell{}H{}$\textsuperscript{E}}},
\mbox{$\ell{}t{}H{}$\textsuperscript{*}},
\mbox{$t{}H{}H{}$\textsuperscript{*}},
\mbox{$t{}Z{}H{}$\textsuperscript{*}},
\mbox{$b{}W{}H{}$\textsuperscript{*}},
\mbox{$j{}W{}H{}$\textsuperscript{*}},
\mbox{$\tau{}b{}H{}$\textsuperscript{*}},
\mbox{$\tau{}j{}H{}$\textsuperscript{*}},
\mbox{$\gamma{}H{}H{}$\textsuperscript{*}},
\mbox{$\gamma{}W{}H{}$\textsuperscript{*}},
\mbox{$\gamma{}t{}H{}$\textsuperscript{*}},
\mbox{$\gamma{}b{}H{}$\textsuperscript{*}},
\mbox{$\gamma{}j{}H{}$\textsuperscript{*}},
\textcolor{red}{\mbox{$b{}H{}H{}$\textsuperscript{S}}},
\textcolor{red}{\mbox{$b{}Z{}H{}$\textsuperscript{S}}},
\textcolor{red}{\mbox{$j{}H{}H{}$\textsuperscript{S}}},
\textcolor{red}{\mbox{$j{}Z{}H{}$\textsuperscript{S}}},
\textcolor{red}{\mbox{$\tau{}\tau{}H{}$\textsuperscript{S}}},
\textcolor{red}{\mbox{$\gamma{}Z{}H{}$\textsuperscript{S}}},
\textcolor{red}{\mbox{$W{}Z{}H{}$\textsuperscript{C}}}.\par
\appfamily{Top (12)}
\mbox{$\gamma{}\ell{}t{}$\textsuperscript{*}},
\mbox{$\ell{}t{}Z{}$\textsuperscript{*}},
\mbox{$\gamma{}t{}V{}$\textsuperscript{*}},
\mbox{$\tau{}t{}W{}$\textsuperscript{*}},
\mbox{$e{}\tau{}t{}$\textsuperscript{*}},
\mbox{$\gamma{}b{}t{}$\textsuperscript{*}},
\textcolor{red}{\mbox{$\tau{}t{}t{}$\textsuperscript{LQ}}}.\par
\appfamily{Vector boson (18)}
\mbox{$\ell{}b{}V{}$\textsuperscript{*}},
\mbox{$\ell{}j{}Z{}$\textsuperscript{*}},
\mbox{$b{}W{}Z{}$\textsuperscript{*}},
\mbox{$\tau{}j{}W{}$\textsuperscript{*}},
\mbox{$\gamma{}\tau{}W{}$\textsuperscript{*}},
\textcolor{red}{\mbox{$\gamma{}\tau{}Z{}$\textsuperscript{E}}},
\textcolor{red}{\mbox{$b{}Z{}Z{}$\textsuperscript{S}}},
\textcolor{red}{\mbox{$j{}Z{}Z{}$\textsuperscript{S}}},
\textcolor{red}{\mbox{$\gamma{}Z{}Z{}$\textsuperscript{S}}},
\textcolor{red}{\mbox{$\gamma{}b{}Z{}$\textsuperscript{S}}},
\textcolor{red}{\mbox{$\gamma{}j{}Z{}$\textsuperscript{S}}}.\par
\appfamily{$\tau$ (4)}
\mbox{$\gamma{}\tau{}b{}$\textsuperscript{*}},
\mbox{$\gamma{}\tau{}j{}$\textsuperscript{*}},
\textcolor{red}{\mbox{$\tau{}b{}b{}$\textsuperscript{LQ}}},
\textcolor{red}{\mbox{$\tau{}j{}b{}$\textsuperscript{LQ}}}.\par
\appfamily{Photon (1)}
\textcolor{red}{\mbox{$\gamma{}\gamma{}b{}$\textsuperscript{S}}}.\par
\appheading{Exactly 4 visible objects (507 signatures)}
\appfamily{Higgs (183)}
\mbox{$\ell{}V{}V{}H{}$\textsuperscript{*}},
\mbox{$\gamma{}\ell{}\ell{}H{}$\textsuperscript{*}},
\mbox{$\gamma{}\ell{}V{}H{}$\textsuperscript{*}},
\mbox{$\ell{}b{}V{}H{}$\textsuperscript{*}},
\mbox{$\ell{}j{}V{}H{}$\textsuperscript{*}},
\mbox{$\ell{}\ell{}b{}H{}$\textsuperscript{*}},
\mbox{$\ell{}\ell{}j{}H{}$\textsuperscript{*}},
\mbox{$\ell{}\ell{}t{}H{}$\textsuperscript{*}},
\mbox{$\ell{}t{}V{}H{}$\textsuperscript{*}},
\mbox{$\ell{}V{}H{}H{}$\textsuperscript{*}},
\mbox{$b{}V{}V{}H{}$\textsuperscript{*}},
\mbox{$\gamma{}\gamma{}\ell{}H{}$\textsuperscript{*}},
\mbox{$\gamma{}\ell{}H{}H{}$\textsuperscript{*}},
\mbox{$\gamma{}\ell{}b{}H{}$\textsuperscript{*}},
\mbox{$\gamma{}\ell{}j{}H{}$\textsuperscript{*}},
\mbox{$\gamma{}\ell{}t{}H{}$\textsuperscript{*}},
\mbox{$j{}V{}V{}H{}$\textsuperscript{*}},
\mbox{$\ell{}H{}H{}H{}$\textsuperscript{*}},
\mbox{$\ell{}b{}H{}H{}$\textsuperscript{*}},
\mbox{$\ell{}b{}b{}H{}$\textsuperscript{*}},
\mbox{$\ell{}b{}t{}H{}$\textsuperscript{*}},
\mbox{$\ell{}j{}H{}H{}$\textsuperscript{*}},
\mbox{$\ell{}j{}b{}H{}$\textsuperscript{*}},
\mbox{$\ell{}j{}j{}H{}$\textsuperscript{*}},
\mbox{$\ell{}j{}t{}H{}$\textsuperscript{*}},
\mbox{$\ell{}t{}H{}H{}$\textsuperscript{*}},
\mbox{$\ell{}t{}t{}H{}$\textsuperscript{*}},
\mbox{$t{}V{}V{}H{}$\textsuperscript{*}},
\mbox{$b{}V{}H{}H{}$\textsuperscript{*}},
\mbox{$\gamma{}W{}V{}H{}$\textsuperscript{*}},
\mbox{$\gamma{}b{}V{}H{}$\textsuperscript{*}},
\mbox{$\gamma{}j{}V{}H{}$\textsuperscript{*}},
\mbox{$\gamma{}t{}V{}H{}$\textsuperscript{*}},
\mbox{$j{}V{}H{}H{}$\textsuperscript{*}},
\mbox{$t{}V{}H{}H{}$\textsuperscript{*}},
\mbox{$W{}H{}H{}H{}$\textsuperscript{*}},
\mbox{$W{}Z{}H{}H{}$\textsuperscript{*}},
\mbox{$W{}Z{}Z{}H{}$\textsuperscript{*}},
\mbox{$W{}W{}W{}H{}$\textsuperscript{*}},
\mbox{$t{}H{}H{}H{}$\textsuperscript{*}},
\mbox{$t{}t{}W{}H{}$\textsuperscript{*}},
\mbox{$t{}t{}t{}H{}$\textsuperscript{*}},
\mbox{$b{}H{}H{}H{}$\textsuperscript{*}},
\mbox{$b{}t{}H{}H{}$\textsuperscript{*}},
\mbox{$b{}t{}t{}H{}$\textsuperscript{*}},
\mbox{$b{}b{}W{}H{}$\textsuperscript{*}},
\mbox{$b{}b{}t{}H{}$\textsuperscript{*}},
\mbox{$b{}b{}b{}H{}$\textsuperscript{*}},
\mbox{$j{}H{}H{}H{}$\textsuperscript{*}},
\mbox{$j{}t{}H{}H{}$\textsuperscript{*}},
\mbox{$j{}t{}t{}H{}$\textsuperscript{*}},
\mbox{$j{}b{}b{}H{}$\textsuperscript{*}},
\mbox{$e{}\mu{}\tau{}H{}$\textsuperscript{*}},
\mbox{$\gamma{}H{}H{}H{}$\textsuperscript{*}},
\mbox{$\gamma{}W{}H{}H{}$\textsuperscript{*}},
\mbox{$\gamma{}t{}H{}H{}$\textsuperscript{*}},
\mbox{$\gamma{}b{}H{}H{}$\textsuperscript{*}},
\mbox{$\gamma{}b{}t{}H{}$\textsuperscript{*}},
\mbox{$\gamma{}j{}H{}H{}$\textsuperscript{*}},
\mbox{$\gamma{}j{}t{}H{}$\textsuperscript{*}},
\mbox{$\gamma{}\gamma{}W{}H{}$\textsuperscript{*}},
\mbox{$\gamma{}\gamma{}t{}H{}$\textsuperscript{*}},
\mbox{$\gamma{}\gamma{}b{}H{}$\textsuperscript{*}},
\mbox{$\gamma{}\gamma{}j{}H{}$\textsuperscript{*}},
\mbox{$\gamma{}\gamma{}\gamma{}H{}$\textsuperscript{*}},
\textcolor{red}{\mbox{$H{}H{}H{}H{}$\textsuperscript{S}}},
\textcolor{red}{\mbox{$Z{}H{}H{}H{}$\textsuperscript{S}}},
\textcolor{red}{\mbox{$Z{}Z{}H{}H{}$\textsuperscript{S}}},
\textcolor{red}{\mbox{$Z{}Z{}Z{}H{}$\textsuperscript{S}}},
\textcolor{red}{\mbox{$W{}W{}H{}H{}$\textsuperscript{S}}},
\textcolor{red}{\mbox{$W{}W{}Z{}H{}$\textsuperscript{S}}},
\textcolor{red}{\mbox{$\gamma{}Z{}H{}H{}$\textsuperscript{S}}},
\textcolor{red}{\mbox{$\gamma{}Z{}Z{}H{}$\textsuperscript{S}}},
\textcolor{red}{\mbox{$\gamma{}\gamma{}H{}H{}$\textsuperscript{S}}},
\textcolor{red}{\mbox{$\gamma{}\gamma{}Z{}H{}$\textsuperscript{S}}},
\textcolor{red}{\mbox{$\tau{}\tau{}\ell{}H{}$\textsuperscript{E}}},
\textcolor{red}{\mbox{$e{}e{}\ell{}H{}$\textsuperscript{E}}},
\textcolor{red}{\mbox{$\ell{}\mu{}\mu{}H{}$\textsuperscript{E}}},
\textcolor{red}{\mbox{$\gamma{}t{}t{}H{}$\textsuperscript{E}}},
\textcolor{red}{\mbox{$\gamma{}b{}b{}H{}$\textsuperscript{E}}},
\textcolor{red}{\mbox{$\gamma{}j{}b{}H{}$\textsuperscript{E}}}.\par
\appfamily{Top (129)}
\mbox{$\ell{}\ell{}\ell{}t{}$\textsuperscript{*}},
\mbox{$\ell{}t{}V{}V{}$\textsuperscript{*}},
\mbox{$e{}\ell{}t{}V{}$\textsuperscript{*}},
\mbox{$\gamma{}\ell{}\ell{}t{}$\textsuperscript{*}},
\mbox{$\gamma{}\ell{}t{}V{}$\textsuperscript{*}},
\mbox{$\ell{}b{}t{}V{}$\textsuperscript{*}},
\mbox{$\ell{}j{}t{}V{}$\textsuperscript{*}},
\mbox{$\ell{}t{}t{}V{}$\textsuperscript{*}},
\mbox{$t{}V{}V{}V{}$\textsuperscript{*}},
\mbox{$\gamma{}\gamma{}\ell{}t{}$\textsuperscript{*}},
\mbox{$\gamma{}\ell{}b{}t{}$\textsuperscript{*}},
\mbox{$\gamma{}\ell{}j{}t{}$\textsuperscript{*}},
\mbox{$\gamma{}\ell{}t{}t{}$\textsuperscript{*}},
\mbox{$\gamma{}t{}V{}V{}$\textsuperscript{*}},
\mbox{$\ell{}b{}b{}t{}$\textsuperscript{*}},
\mbox{$\ell{}b{}t{}t{}$\textsuperscript{*}},
\mbox{$\ell{}j{}b{}t{}$\textsuperscript{*}},
\mbox{$\ell{}j{}j{}t{}$\textsuperscript{*}},
\mbox{$\ell{}j{}t{}t{}$\textsuperscript{*}},
\mbox{$\ell{}t{}t{}t{}$\textsuperscript{*}},
\mbox{$b{}t{}t{}V{}$\textsuperscript{*}},
\mbox{$\gamma{}b{}t{}V{}$\textsuperscript{*}},
\mbox{$\gamma{}\gamma{}t{}V{}$\textsuperscript{*}},
\mbox{$\gamma{}j{}t{}V{}$\textsuperscript{*}},
\mbox{$\mu{}\tau{}t{}V{}$\textsuperscript{*}},
\mbox{$\mu{}\mu{}t{}V{}$\textsuperscript{*}},
\mbox{$t{}t{}W{}Z{}$\textsuperscript{*}},
\mbox{$t{}t{}t{}Z{}$\textsuperscript{*}},
\mbox{$b{}t{}W{}W{}$\textsuperscript{*}},
\mbox{$b{}b{}t{}Z{}$\textsuperscript{*}},
\mbox{$j{}t{}t{}Z{}$\textsuperscript{*}},
\mbox{$\tau{}\tau{}t{}Z{}$\textsuperscript{*}},
\mbox{$\mu{}\tau{}t{}t{}$\textsuperscript{*}},
\mbox{$\mu{}\tau{}b{}t{}$\textsuperscript{*}},
\mbox{$e{}\tau{}t{}t{}$\textsuperscript{*}},
\mbox{$e{}\tau{}b{}t{}$\textsuperscript{*}},
\mbox{$e{}\tau{}j{}t{}$\textsuperscript{*}},
\mbox{$e{}\mu{}t{}t{}$\textsuperscript{*}},
\mbox{$e{}\mu{}b{}t{}$\textsuperscript{*}},
\mbox{$\gamma{}t{}t{}W{}$\textsuperscript{*}},
\mbox{$\gamma{}t{}t{}t{}$\textsuperscript{*}},
\mbox{$\gamma{}b{}t{}t{}$\textsuperscript{*}},
\mbox{$\gamma{}b{}b{}t{}$\textsuperscript{*}},
\mbox{$\gamma{}j{}b{}t{}$\textsuperscript{*}},
\mbox{$\gamma{}j{}j{}t{}$\textsuperscript{*}},
\mbox{$\gamma{}\gamma{}b{}t{}$\textsuperscript{*}},
\mbox{$\gamma{}\gamma{}\gamma{}t{}$\textsuperscript{*}},
\textcolor{red}{\mbox{$\gamma{}t{}t{}Z{}$\textsuperscript{S}}},
\textcolor{red}{\mbox{$\tau{}\tau{}b{}t{}$\textsuperscript{LQ}}},
\textcolor{red}{\mbox{$\mu{}\mu{}b{}t{}$\textsuperscript{LQ}}},
\textcolor{red}{\mbox{$e{}e{}b{}t{}$\textsuperscript{LQ}}}.\par
\appfamily{Vector boson (154)}
\mbox{$\ell{}\ell{}b{}V{}$\textsuperscript{*}},
\mbox{$\ell{}V{}V{}V{}$\textsuperscript{*}},
\mbox{$\ell{}\ell{}\ell{}W{}$\textsuperscript{*}},
\mbox{$\gamma{}\ell{}V{}V{}$\textsuperscript{*}},
\mbox{$\ell{}b{}V{}V{}$\textsuperscript{*}},
\mbox{$\ell{}j{}V{}V{}$\textsuperscript{*}},
\mbox{$\tau{}\ell{}j{}V{}$\textsuperscript{*}},
\mbox{$\gamma{}\gamma{}\ell{}V{}$\textsuperscript{*}},
\mbox{$\gamma{}\ell{}b{}V{}$\textsuperscript{*}},
\mbox{$\gamma{}\ell{}j{}V{}$\textsuperscript{*}},
\mbox{$\gamma{}\ell{}\ell{}W{}$\textsuperscript{*}},
\mbox{$b{}V{}V{}V{}$\textsuperscript{*}},
\mbox{$j{}V{}V{}V{}$\textsuperscript{*}},
\mbox{$\gamma{}b{}V{}V{}$\textsuperscript{*}},
\mbox{$\gamma{}j{}V{}V{}$\textsuperscript{*}},
\mbox{$\ell{}b{}b{}Z{}$\textsuperscript{*}},
\mbox{$\ell{}j{}b{}Z{}$\textsuperscript{*}},
\mbox{$b{}b{}b{}V{}$\textsuperscript{*}},
\mbox{$e{}\mu{}j{}V{}$\textsuperscript{*}},
\mbox{$\gamma{}\gamma{}b{}V{}$\textsuperscript{*}},
\mbox{$\gamma{}\gamma{}j{}V{}$\textsuperscript{*}},
\mbox{$W{}Z{}Z{}Z{}$\textsuperscript{*}},
\mbox{$b{}b{}W{}Z{}$\textsuperscript{*}},
\mbox{$\tau{}j{}j{}Z{}$\textsuperscript{*}},
\mbox{$\mu{}\tau{}W{}W{}$\textsuperscript{*}},
\mbox{$\mu{}\mu{}j{}Z{}$\textsuperscript{*}},
\mbox{$e{}\tau{}W{}W{}$\textsuperscript{*}},
\mbox{$e{}\mu{}W{}W{}$\textsuperscript{*}},
\mbox{$e{}\mu{}\tau{}Z{}$\textsuperscript{*}},
\mbox{$e{}e{}j{}Z{}$\textsuperscript{*}},
\mbox{$\gamma{}W{}Z{}Z{}$\textsuperscript{*}},
\mbox{$\gamma{}W{}W{}W{}$\textsuperscript{*}},
\mbox{$\gamma{}b{}b{}W{}$\textsuperscript{*}},
\mbox{$\gamma{}j{}b{}W{}$\textsuperscript{*}},
\mbox{$\gamma{}\mu{}\tau{}Z{}$\textsuperscript{*}},
\mbox{$\gamma{}e{}\tau{}Z{}$\textsuperscript{*}},
\mbox{$\gamma{}e{}\mu{}Z{}$\textsuperscript{*}},
\mbox{$\gamma{}\gamma{}W{}Z{}$\textsuperscript{*}},
\mbox{$\gamma{}\gamma{}\gamma{}W{}$\textsuperscript{*}},
\textcolor{red}{\mbox{$\gamma{}Z{}Z{}Z{}$\textsuperscript{S}}},
\textcolor{red}{\mbox{$\gamma{}W{}W{}Z{}$\textsuperscript{S}}},
\textcolor{red}{\mbox{$\gamma{}b{}b{}Z{}$\textsuperscript{S}}},
\textcolor{red}{\mbox{$\gamma{}j{}b{}Z{}$\textsuperscript{S}}},
\textcolor{red}{\mbox{$\gamma{}\tau{}\tau{}Z{}$\textsuperscript{S}}},
\textcolor{red}{\mbox{$\gamma{}\mu{}\mu{}Z{}$\textsuperscript{S}}},
\textcolor{red}{\mbox{$\gamma{}e{}e{}Z{}$\textsuperscript{S}}},
\textcolor{red}{\mbox{$\gamma{}\gamma{}\gamma{}Z{}$\textsuperscript{S}}},
\textcolor{red}{\mbox{$\tau{}\tau{}\ell{}Z{}$\textsuperscript{E}}},
\textcolor{red}{\mbox{$e{}e{}\ell{}Z{}$\textsuperscript{E}}},
\textcolor{red}{\mbox{$\ell{}\mu{}\mu{}Z{}$\textsuperscript{E}}}.\par
\appfamily{$\tau$ (36)}
\mbox{$\tau{}\ell{}\ell{}b{}$\textsuperscript{*}},
\mbox{$\tau{}\ell{}\ell{}j{}$\textsuperscript{*}},
\mbox{$\gamma{}\tau{}\ell{}\ell{}$\textsuperscript{*}},
\mbox{$\gamma{}\tau{}\ell{}b{}$\textsuperscript{*}},
\mbox{$\gamma{}\tau{}\ell{}j{}$\textsuperscript{*}},
\mbox{$\tau{}b{}b{}b{}$\textsuperscript{*}},
\mbox{$\tau{}j{}b{}b{}$\textsuperscript{*}},
\mbox{$\tau{}j{}j{}b{}$\textsuperscript{*}},
\mbox{$\tau{}j{}j{}j{}$\textsuperscript{*}},
\mbox{$\gamma{}\tau{}b{}b{}$\textsuperscript{*}},
\mbox{$\gamma{}\tau{}j{}b{}$\textsuperscript{*}},
\mbox{$\gamma{}\tau{}j{}j{}$\textsuperscript{*}},
\mbox{$\gamma{}\gamma{}\tau{}b{}$\textsuperscript{*}},
\mbox{$\gamma{}\gamma{}\tau{}j{}$\textsuperscript{*}},
\mbox{$\gamma{}\gamma{}\mu{}\tau{}$\textsuperscript{*}},
\mbox{$\gamma{}\gamma{}e{}\tau{}$\textsuperscript{*}},
\mbox{$\gamma{}\gamma{}\gamma{}\tau{}$\textsuperscript{*}}.\par
\appfamily{Photon (5)}
\mbox{$\gamma{}b{}b{}b{}$\textsuperscript{*}},
\mbox{$\gamma{}\gamma{}\gamma{}b{}$\textsuperscript{*}},
\mbox{$\gamma{}\gamma{}\gamma{}j{}$\textsuperscript{*}},
\textcolor{red}{\mbox{$\gamma{}j{}b{}b{}$\textsuperscript{F}}},
\textcolor{red}{\mbox{$\gamma{}j{}j{}b{}$\textsuperscript{F}}}.\par
\end{multicols}

\end{document}